\documentclass{article}      
\usepackage{graphicx}
\usepackage{epstopdf}
\title{Systematic study of
longitudinal
and transverse helicity amplitudes
in the
hypercentral Constituent Quark Model}  

\author{M.M. Giannini\\
Dipartimento di Fisica dell'Universit\`a di Genova\\
}

\date{}
\begin{document}

\maketitle  

\begin{abstract}
  The predictions of the hypercentral Consituent Quark Model for the nucleon helicity amplitudes are briefly reported. Some future perspectives are also discussed.
\end{abstract}

\section{Introduction}
\label{intro}
The hypercentral Constituent Quark Model (hCQM) has been introduced \cite{pl} in order to describe the internal structure of baryons and their electromagnetic properties. After having fixed the few free parameters by fitting the observed spectrum of baryon resonances, the model has been used to predict many baryon properties, such as the photocouplings \cite{aie}, the helicity amplitudes for the transverse excitation of negative parity resonances \cite{aie2} and the nucleon elastic form factors \cite{mds2,rap}. Recently the predicted transverse and longitudinal helicity amplitudes up to $Q^2= 5 (GeV/c)^2$ for both proton and neutron have been published \cite{sg}. In this paper, after a brief description of the model, the attention will be concentrated on the results regarding the excitation of some of the most important resonances and then future perspectives, mainly in connection with the inclusion of relativistic effects, will be discussed. 
\section{The model}
\label{sec-1}
n the hCQM the  three quark hamiltonian is assumed to be \cite{pl}
\begin{equation}\label{hCQM}
H~=~\frac{p_{\lambda}^2}{2m}~+~\frac{p_{\rho}^2}{2m}~ -\frac{\tau}{x}~+~\alpha x~+~H_{hyp}
\end{equation}
where $\vec{p_{\rho}}$ and $\vec{p_{\lambda}}$ are the momenta conjugated to the coordinates $\vec{\rho}~=~ \frac{1}{\sqrt{2}} (\vec{r}_1 - \vec{r}_2) ,
\vec{\lambda}~=~\frac{1}{\sqrt{6}} (\vec{r}_1 + \vec{r}_2 - 2\vec{r}_3)$,  
$H_{hyp}$ is the hyperfine interaction \cite{is} and x is the hyperradius, defined as $x=\sqrt{{\vec{\rho}}^2+{\vec{\lambda}}^2}$.

The spin independent part is assumed to depend on the hyperradius $x$ only, that is to be hypercentral. This assumption is supported by the fact that two-body potentials, when applied to baryons, behave approximately as a hypercentral one  \cite{has}. In this sense the  coulomblike and the  linear confining term  in Eq. (\ref{hCQM}) can be considered as the hypercenral approximation of a two-body potential of the form suggested by lattice 
QCD calculations \cite{bali}.
On the other hand, a hypercentral potential may contain contributions from three-body forces, which have been taken into account also in the calculations by ref. \cite{ckp}
and in the relativized version of the Isgur-Karl model \cite{ci}.

Having fixed the quark mass $m$ to $1/3$ of the nucleon mass and the strength of the 
hyperfine interaction in order to reproduce the $\Delta$ - Nucleon mass difference,   
the remaining parameters are fitted to the spectrum and are given by $\tau=4.59$
and $\alpha=1.61~fm^{-2}$ \cite{pl}. The three-quark wave functions calculated with the hCQM of Eq. (\ref{hCQM}) can be used for the prediction of 
the $Q^2$ behavior of the helicity amplitudes \cite{aie,aie2,sg} and of the nucleon elastic form factors \cite{mds2,rap}.

\begin{figure}
\centering
\includegraphics[width=7cm]{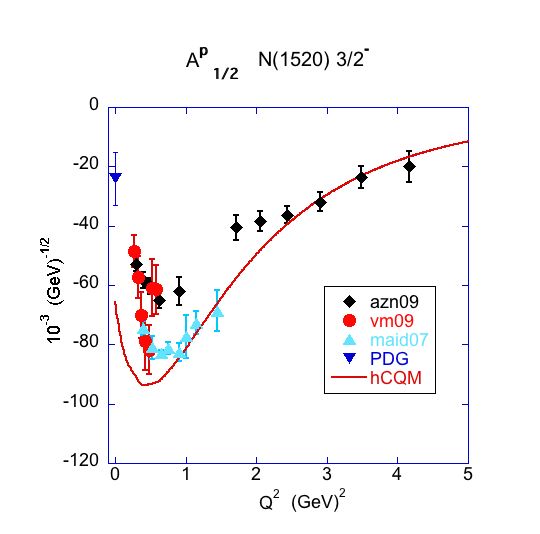}
\includegraphics[width=7cm]{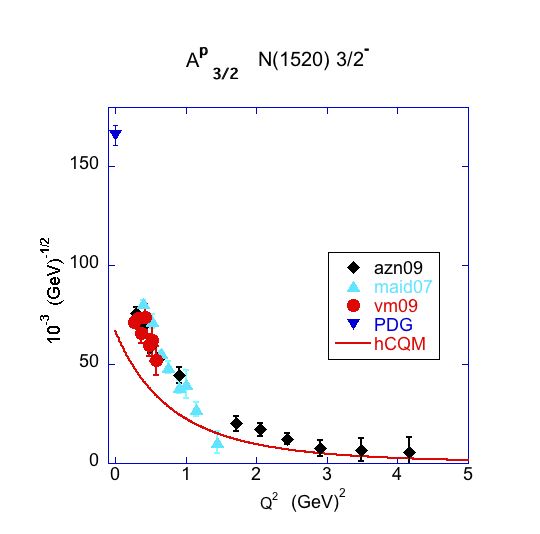}
\caption{The proton helicity amplitudes for the excitation to the N(1520) resonance: $A^p_{3/2}$ (left), $A^p_{1/2}$ (right). The data are taken from \cite{azn09} (diamonds), \cite{vm09} (full points), \cite{pdg} (downward triangles), \cite{maid07} (upward triangles).}  
\label{fig-1}      
\end{figure}

\section{The helicity amplitudes}
\label{sec-2}

The electromagnetic transition amplitudes, 
$A_{1/2}$, $A_{3/2}$ and $S_{1/2}$, are defined as the matrix elements of
the quark electromagnetic interaction, $A_{\mu} J^{\mu}$, between the nucleon, 
$N$, and the resonance, $B$, states:
\begin{equation} \label{eq:hel}
\begin{array}{rcl}
A_{M}&=&   \sqrt{\frac{2 \pi \alpha}{k}}   \langle B, J', J'_{z}=M\ | J_{+}| N, J~=~
\frac{1}{2}, J_{z}= M-1 \
\rangle , ~~~~~~~~~~M=\frac{1}{2}, \frac{3}{2}\\
& & \\
S_{1/2}&=&   \sqrt{\frac{2 \pi \alpha}{k}}   \langle B, J', J'_{z}=\frac{1}{2}\ | J_{z}| N, J~=~
\frac{1}{2}, J_{z}= -\frac{1}{2}\
\rangle\\\end{array}
\end{equation}
$J_{\mu}$ is the electromagnetic current carried by quarks and will be used in its non relativistic  form 
\cite{cko, ki}; $k$ is the photon momentum in the Breit frame.

The results for the photocouplings, that is the transverse amplitudes with zero photon tetramomentum ($Q^2=0$)  compare favourably with other model \cite{aie}. The overall trend is reproduced, but as in all models, there is  a lack of strength.

A systematic calculation of the helicity amplitudes for the transverse and longitudinal excitation to 14 resonances of both proton and neutron has been recently reported \cite{sg}. For many resonances there are still few data, however, thanks to the recent Jlab experiments at higher $Q^2$, one can make a significant comparison between the extracted amplitudes and the hCQM predictions. As an example, in Fig. (\ref{fig-1}) the transverse helicity amplitudes for the proton excitation to the  N(1520) $1/2^-$ resonance are given.  It should be mentioned that the predictions for the transverse excitation to the negative parity states had been published \cite{aie2} before the appearance of the new Jlab data. 

Apart from the low $Q^2$ region, where there is the already quoted missing strength, the overall behaviour is in general fairly reproduced, specially for the 1/2 amplitudes. The discrepancy at low $Q^2$ is attributed to the missing quark-antiquark pair or meson production mechanisms \cite{aie2}. Actually, the calculation of the meson cloud effects in the DMT model \cite{dmt} show that their contribution is relevant for low $Q^2$, just in the region where the hCQM fails to reproduce the strength \cite{lt}.

The data mainly concern the proton excitation, however there are now analyses which succeed in extracting the neutron photocouplings \cite{anis}. In Table \ref{tab-1} a comparison between the Bonn data \cite{anis} and the predicton of the hCQM \cite{sg} for the  A1/2 neutron photocoupling is given.

\begin{table}
\centering
\caption{The neutron photocouplings to various resonances. Data labeled as Bonn are taken from \cite{anis}, while the theoretical predictions of the hCQM are from \cite{sg}. (Units in $10^{-3} (GeV)^{-1/2}$)}
\label{tab-1}       
\begin{tabular}{|c|c|c||c|c|c|}
\hline
State & Bonn & hCQM & State & Bonn & hCQM \\\hline
N(1440) & $43 \pm 12$ & 58  & N(1675) & $-60 \pm 7 $ & -37  \\\hline
N(1520) & $-49 \pm 8$ & -1 & N(1680) & $ 34 \pm 6 $ &  38 \\\hline
N(1535) & $-93 \pm 11 $ & -82 & N(1710) & $-40 \pm 20 $ & -22 \\\hline
N(1650) & $ 25 \pm 20 $ & -21 & N(1720) & $-80 \pm 50$ & -48 \\\hline
\end{tabular}
\end{table}

\section{Relativity}
\label{sec-3}

It is possible to introduce relativistic corrections to the hCQM calculations for the nucleon elastic form factors \cite{mds} and the helicity ampltudes \cite{mds2}. The relativistic corrections modify only slightly the helicity amplitudes \cite{mds2}, apart from the case of the $\Delta$ excitation, and do not explain the missing strength at low $Q^2$. On the contrary, they are very important for the elastic form factors \cite{mds}.

For this reason the hCQM has been reformulated in a relativistically consistent framework using the Point form formulation of the Dirac relativistic dynamics \cite{dirac} and has been applied to the calculation of the elastic nucleon form factors \cite{ff_07,ff_10} and the predicted values are  very near to the experimental data; a further improvement is obtained  if quark form factors are introduced \cite{ff_07,ff_10}, in particular for the ratio $\mu_p G_E(Q^2)/G_M(Q^2)$.

For consistency, a programme for the calculation of the helicity amplitudes in the relativistic hCQM has been started, The preliminary results for the $\Delta$ excitation show  that the relativistic are, as expected, very important and bring the predictions quite near to the experimental data \cite{dong_efb22}.

Another important issue in connection with relativity is that of the meson cloud effects. To thos end one has to introduce in a consistent way the quark-antiquark pair creation mechanism into the CQM, that is one gas to unquench the quark model, similarly to what happens in LQCD.

The unquenching of the quark model for the meson sector has been performed long go \cite{ge_is}, but only recently it has been obtained also for the baryon sector \cite{sb1, bs}. This new formulation is an important breakthrough for CQMs, in particular it will allow to take properly into account the coupling with the continuum and predict a non zero width for the baryon resonances. It has been shown that the unquenched CQM (U\cite{}CQM) does not modify 
the old good results for the magnetic moments and that it gives important results for the spin proton problem, the flavour asymmetry of sea quarks \cite{sb2} and the problem if the strangeness content of baryons \cite{sbf,bfs}. 

 To conclude, the hCQM is a useful tool for a consistent description of various electromagnetic properties and may also be helpful as a support to the planning of future experiments at the 12 GeV upgrade of Jefferson Lab \cite{azn}.

\end{document}